\documentclass[conference]{IEEEtran}
\usepackage[linesnumbered,lined,boxed,commentsnumbered]{algorithm2e}
\usepackage{amsmath,amssymb,amsfonts}
\usepackage{algorithmic}
\usepackage{graphicx}
\usepackage{textcomp}
\usepackage{xcolor}
\usepackage{amsthm}
\theoremstyle{definition}

\usepackage{array}
\usepackage{bm}
\usepackage{booktabs}
\usepackage{cite}
\usepackage{cleveref}
\usepackage{color}
\usepackage{dblfloatfix}
\usepackage{enumitem}
\usepackage{graphicx}
\usepackage{scalerel}

\usepackage{soul}
\usepackage[caption=false,font=footnotesize]{subfig}
\usepackage{url}

\def\BibTeX{{\rm B\kern-.05em{\sc i\kern-.025em b}\kern-.08em
    T\kern-.1667em\lower.7ex\hbox{E}\kern-.125emX}}

\newcommand{\aperture}{\mathcal{A}}

\begin{document}

\title{Minimum Norm Method for Linear and Planar Sparse Arrays}

\author{\IEEEauthorblockN{Tyler M. Trosclair, Kaushallya Adhikari}
\IEEEauthorblockA{\textit{University of Rhode Island, Kingston, RI, USA} \\
\{tyler\_trosclair,kadhikari\}@uri.edu}
}

\maketitle

\begin{abstract}
Coprime and nested arrays are sparse  arrays with enhanced degrees of freedom, which  can be exploited in direction of arrival estimation using algorithms such as product processing, min processing, and MUSIC. This paper applies the minimum norm method for direction of arrival estimation. Comparison of the root mean squared errors and probabilities of resolution  of the minimum norm method with MUSIC for a given linear coprime or nested array demonstrates the superiority of the minimum norm method. Specifically, minimum norm method exhibits lower mean squared error, narrower peaks at the locations of the true sources, and a lower noise floor in the spatial spectral estimate. This work also formulates two different minimum norm methods for planar sparse arrays: direct and linear. Comparison of the linear minimum norm method with the linear MUSIC for planar arrays also demonstrates higher accuracy of the minimum norm method.
\end{abstract}

\begin{IEEEkeywords}
coprime arrays, DoA estimation, minimum norm method, MUSIC, nested arrays, sparse arrays.
\end{IEEEkeywords}

\section{Introduction}
Direction of arrival (DoA) estimation of a propagating signal using data received by a sensor array is a crucial task in many fields including  acoustics, sonar, radar, radio astronomy, and seismology. A full sensor array has equally spaced sensors with an intersensor spacing of $\lambda/2,$ where $\lambda$ is the wavelength of the propagating signal. The aperture and intersensor spacing are two vital array parameters in DoA estimation.  In general, increasing the array aperture improves the resolution in DoA estimation. As dictated by the Nyquist sampling theorem, when the intersensor spacing exceeds $\lambda/2,$ the DoAs cannot be resolved unambiguously  \cite{VanTrees}. Arrays with an average intersensor spacing larger than $\lambda/2$ are known as sparse arrays. There exist many sparse array designs in which ambiguity caused by the violation of Nyquist sampling theorem can be disambiguated using prudent processing techniques. Coprime arrays and nested arrays are two widely studied examples of such sparse arrays \cite{VandP1}, \cite{nested1}. Both coprime and nested arrays are formed by interleaving two uniform linear arrays (ULAs), henceforth called Subarray~1 and Subarray~2.

The majority of DoA estimation algorithms applicable to coprime and nested arrays fall into two general categories: (1) conventional beamforming-based algorithms and (2) \textit{eigenanalysis} algorithms  \cite{VandP1,nested1,shadings,wagemagazine,adhikariJasa1,chavali2,psdestimation, AdhikariNAECON}. Algorithms that apply conventional beamforming (CBF) to individual subarrays include product processing and min processing \cite{VandP1,nested1,wagemagazine, shadings,asaconf2,psdestimation,asaconf1,detection2019,chavali2,asaconf3,
icasspdetection,liubuck4,adhikariJasa1,liubuck4,AdhikariAccess1}.  Product processing multiplies one subarray's CBF output with the complex conjugate of the other subarray's output to estimate the spatial power spectrum, whereas min processing finds the minimum of the magnitudes of the CBF outputs. Both algorithms have low resolution.  Eigenanalysis algorithms have been widely studied for DoA estimation with coprime and nested arrays \cite{VandP1,nested1 ,AdhikariAccess2}. The most prominent eigenanalysis algorithm that has been applied to coprime and nested arrays is  multiple signal classification (MUSIC) \cite{Schmidt}.  The minimum norm method (MNM) is another eigenanalysis method that has been proven superior to MUSIC for full arrays \cite{KumaresanTufts}. However, MNM has not yet been applied to coprime and nested arrays. In this paper, we formulate MNM for sparse arrays. We also propose an algorithm to integrate MNM to planar coprime and nested arrays. We demonstrate the  efficacy of the MNM-based algorithms in DoA estimation and show how  MNM is more robust than MUSIC for a wide range of signal-to-noise ratios (SNRs) and numbers of snapshots.  The specific contributions of this paper are:
\begin{enumerate}
\item A demonstration that MNM's advantages over MUSIC for standard ULAs  also extends to sparse arrays.
\item A formulation of the two-dimensional MNM algorithm for planar sparse arrays.
\item The introduction of a linear MNM algorithm for planar sparse arrays and an illustration of its higher accuracy over the MUSIC counterpart.
\end{enumerate}

Section \ref{sec:2} and Section \ref{sec:planargeometry} describe MNM for linear sparse arrays and planar sparse arrays, respectively.   Section \ref{sec:results} compares the MNM and MUSIC algorithms for both linear and planar sparse arrays. Section \ref{sec:conclusion} summarizes the paper's contributions.

\textit{Notations:} Boldfaced lowercase math symbols denote vectors and boldfaced uppercase math symbols denote matrices. $\mathbf{x}^T$ denotes transpose of $\mathbf{x}$. $\mathbf{x}^H$ denotes Hermitian (conjugate-transpose) of $\mathbf{x}$. $\otimes$ denotes Kronecker product. $vec(\mathbf{W})$ denotes the vector obtained by stacking the columns of $\mathbf{W}$ in order.

\section{MNM for Linear Sparse Arrays}
\label{sec:2}

\subsection{Linear Sparse Array Review}
\label{subsec:2:1}

\begin{figure}[h]
    \centering
    	\includegraphics[width=0.95\columnwidth, trim = 0cm 0cm 0cm 0cm]{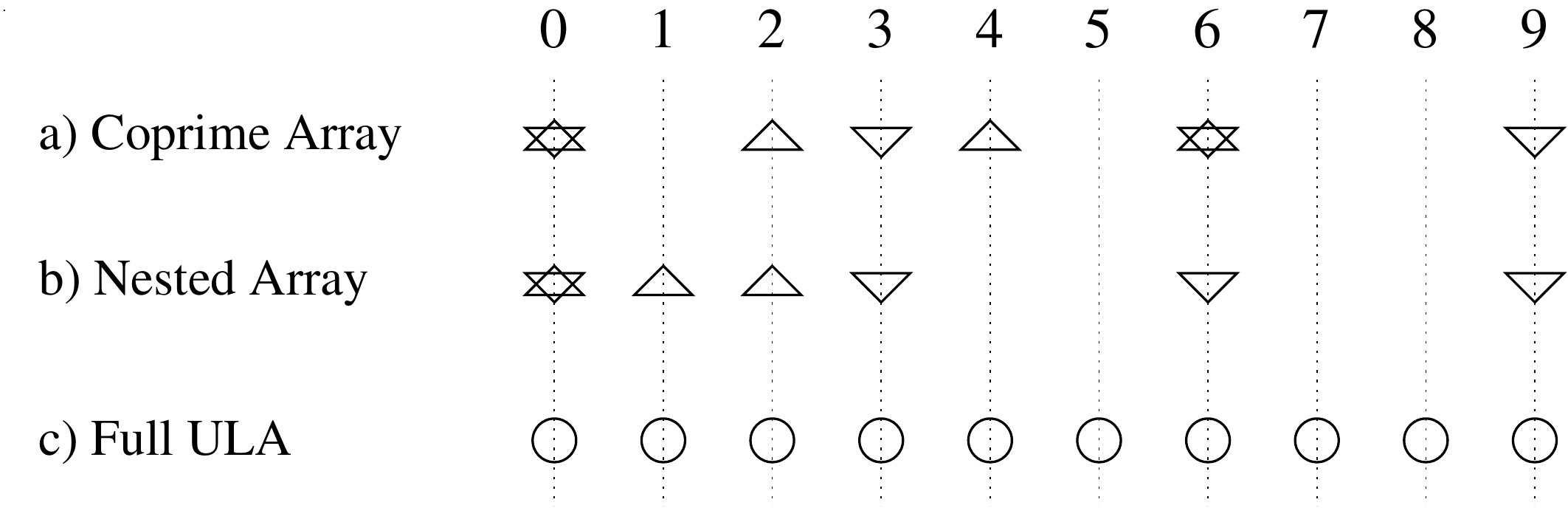}
    
        \caption{Array geometries. Top: A coprime array with parameters $M_e=4$, $N_e=4$, $N=2$, and $M=3$. Middle: A nested array with parameters $M_e=3$, $N_e=4$, $N=1$, and $M=3$. Bottom: A full ULA with $L=10$.\label{arrayslabel}}
        
\end{figure}

We assume that all linear arrays are aligned along the positive $z$-axis and the first sensor is located at $z=0,$ without loss of generality. We focus on a category of  sparse arrays where one sparse array is comprised of two uniform linear subarrays. Coprime and nested arrays fall into this category. The two subarrays are called Subarray 1 and Subarray 2. Subarray 1 consists of $M_e$ sensors and has an undersampling factor of $N.$ Similarly, Subarray 2 consists of $N_e$ sensors and has an undersampling factor of $M.$ In a coprime array, the undersampling factors of Subarray 1 and Subarray 2 are coprime integers. Also, to minimize the total number of sensors, the coprime integers are chosen to satisfy the equation $N=M+1$ \cite{KBW,shadings}. In a nested array, the undersampling factor of Subarray 1 is $N=1$ and the undersampling factor of Subarray~2 is $M=M_e$. The aperture of Subarray 1 and Subarray 2 are $(M_e-1)N\lambda/2$ and $(N_e-1)M\lambda/2,$ respectively, where $\lambda$ is the wavelength of the plane wave signal to be sampled by the array. The aperture of the non-uniform linear array (NULA) formed by interleaving Subarray 1 and Subarray 2 is
\begin{equation}
\nonumber
\aperture=\max((M_e-1)N\lambda/2,(N_e-1)M\lambda/2).
\end{equation} A full ULA that has aperture $\aperture$ consists of $L=1+2\aperture/\lambda$ sensors. Fig. \ref{arrayslabel} depicts a coprime array with parameters $M_e=4$, $N_e=4$, $N=2$, and $M=3$, a nested array with parameters $M_e=3$, $N_e=4$, $N=1$, and $M=3$, and a full ULA with $L=10$ sensors. The three arrays have equal aperture. The coprime and nested arrays also have the same total number of sensors. 

\subsection{Received Signal Model and Correlation Matrix Estimate}
\label{sec:correlation}

We assume that a plane wave impinging on an array has wavelength $\lambda$. The angle of arrival of a plane wave, $\theta_i$,  is measured counterclockwise from the array axis and the goal of the DoA problem is to estimate this angle for every plane wave impinging on the array. We make the following assumptions---both signal and noise are independent zero mean Gaussian random variables, and there are $P$ wide sense stationary (WSS) plane wave signals embedded in additive spatially white noise. The number of available snapshots is denoted by $Q$. The data vector received by an $L$-sensor full ULA corresponding to the $q^{th}$ snapshot is
\begin{equation}
\label{signal1}
\mathbf{x}_q=\sum_{i=1}^P a_{i,q}\mathbf{v}(u_i)+\mathbf{n}_q,
\end{equation}
where $a_{i,q}$ is the $i^{th}$ signal's complex amplitude and  $u_i=\cos(\theta _i)$ is the $i^{th}$ signal's  direction cosine, which represents the DoA. The signal component in \eqref{signal1} is $\sum_{i=1}^P a_{i,q}\mathbf{v}(u_i),$ and the noise component is $\mathbf{n}_q$. The $L$-element vector $\mathbf{v}(u_i)$ is the array manifold vector at direction cosine $u_i=\cos(\theta_i),$  given by $\mathbf{v}(u_i)=\left[\begin{array}{l}1,e^{j\pi u_i},e^{j\pi u_i2},\hdots,e^{j\pi u_i(L-1)}\end{array}\right]^T.$ We define discrete sensor location functions for Subarray~1 and Subarray~2 as $b_1[k]=\sum_{m=0}^{M_e-1}\delta[k-mN]$ and $b_2[k]=\sum_{m=0}^{N_e-1}\delta[k-mM],$ where $\delta[k]$ is the discrete impulse function. For $i=\lbrace 1,2\rbrace$, $b_i[k]=1$ if the $i^{th}$ subarray has a sensor at location $z=k\lambda/2$, and $b_i[k]=0$ otherwise. The sensor location function for the overall NULA obtained by interleaving the two subarrays is $b[k]=b_1[k]\cup b_2[k];$ since $b_i[k]$ is binary, $\cup$ is equivalent to the logical OR operation. When one or both of the subarrays have a sensor at $z=k\lambda/2$, then $b[k]=1$. When neither subarray has a sensor at $z=k\lambda/2$, then $b[k]=0$. The process to obtain the spatial correlation matrix estimate $\mathbf{R}$ corresponding to the received dataset is outlined in detail in \cite{AdhikariNAECON} and \cite{AdhikariAccess2}.

\subsection{Application of MNM to Planar Sparse Arrays}
\label{subsec:mnm}

Let the eigenvalues of $\mathbf{R}$ be $\lambda_1\geq \lambda_2\geq \hdots \geq \lambda_{\mathcal{L}}$, where $\mathcal{L}$ is the rank of $\mathbf{R}.$ Let the corresponding orthonormal eigenvectors of $\mathbf{R}$ be $\mathbf{e}_1,$ $\mathbf{e}_2, \hdots , \mathbf{e}_\mathcal{L}.$ If the measurements consist of $P$ uncorrelated plane waves in spatially white noise, then the correlation matrix can be decomposed as
\begin{equation}
\label{linearR}
\mathbf{R}=\sum_{i=1}^P\mathbf{e}_i\mathbf{e}_i^H\lambda_i+\sigma_n^2\mathbf{I}_{\mathcal{L}},
\end{equation}
where $\sigma_n^2$ is the noise variance at each sensor and $\mathbf{I}_{\mathcal{L}}$ is an $\mathcal{L}$-by-$\mathcal{L}$ identity matrix. Both MUSIC and MNM  take advantage of the fact that signal and noise subspaces of the correlation matrix are orthogonal,
$\mathbf{e}_i^H\mathbf{d}=0$, where $\mathbf{d}$ is any vector in the noise subspace of $\mathbf{R}$. The signal subspace is given by $\mathbf{E}_s=[\mathbf{e}_1,\mathbf{e}_2,\hdots,\mathbf{e}_P]$ and the noise subspace is given by $\mathbf{E}_n=[\mathbf{e}_{P+1},\mathbf{e}_{P+2},\hdots,\mathbf{e}_{\mathcal{L}}].$ The MUSIC algorithm utilizes all the noise eigenvectors (i.e., $\mathbf{e}_i$ for $i=P+1$ to $\mathcal{L}$). The MNM algorithm finds a single vector $\mathbf{d}$ in the noise subspace of the correlation matrix. Moreover, MNM also ensures that the Euclidean norm of $\mathbf{d}$ is minimized, subject to the constraint that its first element equals $1.$ The vector $\mathbf{d}$ obtained in this manner yields more accurate outputs, even at relatively low SNR values, when compared to other eigendecomposition-based methods \cite{KumaresanTufts}.  The expression for $\mathbf{d}$ has been derived in \cite{KumaresanTufts}. Using $\mathbf{d},$ the pseudospectrum estimate is computed as $P_{MN}(u)=(|\mathbf{v}^H(u)\mathbf{d}|)^{-2}.$ This pseudo-spectrum can be compared to the MUSIC pseudospectrum. These MUSIC and MNM algorithms have similar computational complexity, but the MNM algorithm exhibits higher accuracy and better resolution.

\section{MNM for Planar Sparse Arrays}
\label{sec:planargeometry}

In this section, we review the symmetry-imposed planar sparse arrays that were proposed in \cite{AdhikariAccess2}---symmetry-imposed rectangular nested array (SIRNA) and symmetry-imposed rectangular coprime array (SIRCA). Then, we formulate two different MNM-based algorithms applicable to SIRNA and SIRCA.

\subsection{Symmetry-Imposed Planar Sparse Arrays Review}
\label{sec:symmetricarrays}

\begin{figure}[h]
\centering
\includegraphics[width=0.4\textwidth,  trim = 0cm 17cm 0cm 0cm,clip]{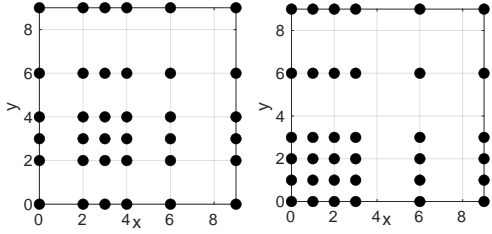}

\caption{Left Panel: Sensor locations of a SIRCA with $M=2$. Right Panel: Sensor locations of a SIRNA with $M=3$ and $N=4$.\label{labelsir}}

\end{figure}

We assume that all planar arrays lie in the $z=0$ plane. Therefore, the sensor locations of a planar array are given by $x$- and $y$-coordinates only. A SIRNA has an underlying linear nested array as defined in Section  \ref{subsec:2:1}, but along the $x$-axis instead of the $z$-axis. The bivariate sensor location function for the underlying linear nested array is $f_L[i,j]=\sum_{m=0}^{M-1}\delta[i-m,j]+\sum_{n=1}^{N-1}\delta[i-Mn,j],$ where $\delta[i,j]$ is the two-dimensional discrete Dirac-delta function. Then, the sensor location function for the SIRNA is given by $f_R[i,j]=\sum_{k\in \boldsymbol{\beta}}f_L[i,j-k]$, where the vector $\boldsymbol{\beta}$ contains the $x$-coordinates of the underlying linear nested array. The value of $f_R[i,j]$ is $1$ when a SIRNA has a sensor at $x=i\lambda/2$ and $y=j\lambda/2$; the value of $f_R[i,j]$ is $0$ when a SIRNA does not  have a sensor at $x=i\lambda/2$ and $y=j\lambda/2$.  The contiguous domain of SIRNA's coarray is $-(N-1)M\leq i,j \leq (N-1)M.$ 

 Similarly, a SIRCA has an underlying linear coprime array with $M_e=2M$ and $N_e=N$ along the $x$-axis. The bivariate sensor location function for the underlying linear coprime array is $f_L[i,j]=\sum_{m=0}^{2M-1}\delta[i-Nm,j]+\sum_{n=1}^{N-1}\delta[i-Mn,j].$ Then, the sensor location function for the SIRCA is given by $f_R[i,j]=\sum_{k\in \boldsymbol{\beta}}f_L[i,j-k]$, where the vector $\boldsymbol{\beta}$ contains the $x$-coordinates of the underlying linear coprime array. The contiguous domain of SIRCA's coarray is $-(MN+M-1)\leq i,j\leq MN+M-1.$ The left and right panels of Fig. \ref{labelsir} illustrate the sensor locations of a SIRCA and a SIRNA. The SIRNA has $M=3$ and $N=4$, and the SIRCA has $M=2$. The total number of sensors for each array is $36$. 
 
\subsection{Received Signal Model and Correlation Matrix Estimate}
\label{sec:planarcorrelation}

\begin{figure}[h]
    \centering
    \includegraphics[width=0.3\textwidth,trim = 0cm 0cm 0cm 0cm ]{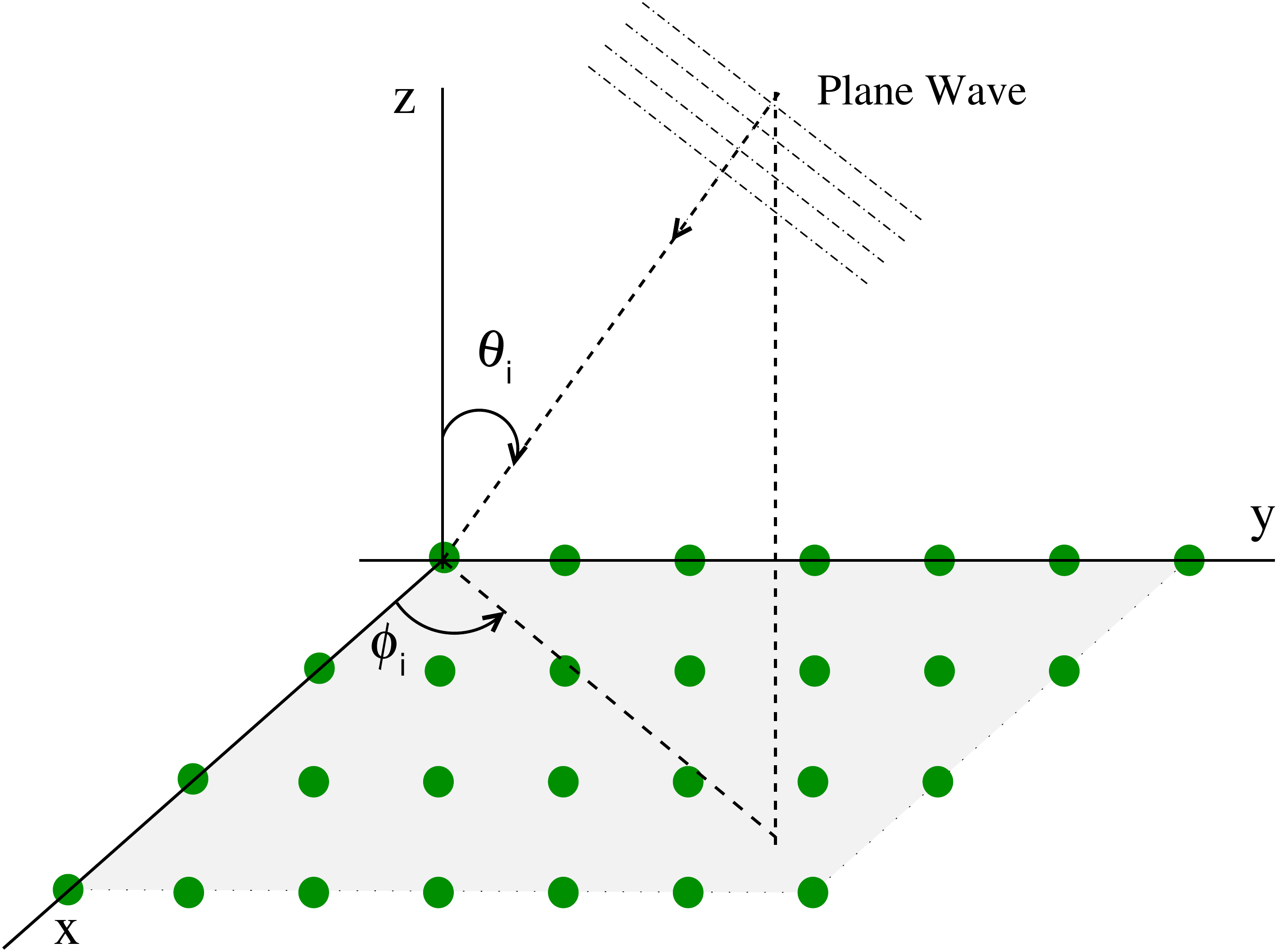}
    
    \caption{A plane wave signal impinging on a planar array making a polar angle of $\theta_i$ with the $z$-axis and an azimuth angle of $\phi_i$ with the $x$-axis.\label{planararraylabel}}

\end{figure}

The DoA of an $i^{th}$ plane wave impinging on a planar array is represented by polar angle $\theta_i$ and azimuth angle $\phi_i$, as depicted in Fig. \ref{planararraylabel}. The polar angle is measured from the $z$-axis to the direction of the plane wave. The azimuth angle is measured from the $x$-axis to the projection of the plane wave's direction onto the $z=0$ plane. The DoA is also completely characterized by the direction cosines pair $u_{x,i}=\sin(\theta_i)\cos(\phi_i)$ and $u_{y,i}=\sin(\theta_i)\sin(\phi_i)$.  We denote $\tilde{\boldsymbol{X}}_q$ as the data matrix received by a planar array that spans from $x=0$ to $(L_x-1)\lambda/2$ and from $y=0$ to $(L_y-1)\lambda/2$. The subscript $q$ represents the snapshot number. The  element at column $i$, row $j$ of $\tilde{\boldsymbol{X}}_q$ contains the data received by the sensor at $x=(i-1)\lambda/2$ and $y=(j-1)\lambda/2$ if a sensor is located at that point. If a sensor is missing at that point, the corresponding element in the data matrix is zero. We represent the dataset  as the vector $\tilde{\boldsymbol{x}}_q=vec(\tilde{\boldsymbol{X}}_q)$, which is given by
\begin{equation}
\label{signal2}
\tilde{\mathbf{x}}_q=\sum_{i=1}^P \tilde{a}_{i,q}\tilde{\mathbf{v}}(u_{x,i},u_{y,i})+\tilde{\mathbf{n}}_q,
\end{equation}
where
\begin{enumerate}
\item $\tilde{a}_{i,q}$ is the $i^{th}$ signal's complex amplitude at snapshot $q$,
\item $\tilde{\boldsymbol{n}}_q=vec(\tilde{\boldsymbol{N}}_q)$, where $\tilde{\boldsymbol{N}}_q$ is the noise matrix with column $i$, row $j$  element containing the noise received by the sensor at $x=(i-1)\lambda/2$ and $y=(j-1)\lambda/2,$
\item $\tilde{\mathbf{v}}(u_{x,i},u_{y,i})$ is the steering vector for direction $(u_{x,i},u_{y,i})$.
\end{enumerate}
The steering vector is given by $\tilde{\mathbf{v}}(u_{x,i},u_{y,i})=\mathbf{w}_{x,i}\otimes \mathbf{w}_{y,i}$, where $\mathbf{w}_{x,i} = \left[\begin{array}{l}1,e^{j\pi u_{x,i}},e^{j\pi u_{x,i}2},\hdots,e^{j\pi u_{x,i}(L_x-1)}\end{array}\right]^T$ and $\mathbf{w}_{y,i} = \left[\begin{array}{l}1,e^{j\pi u_{y,i}},e^{j\pi u_{y,i}2},\hdots,e^{j\pi u_{y,i}(L_y-1)}\end{array}\right]^T.$

The procedure to obtain a block diagonal unbiased correlation matrix estimate, $\tilde{\mathbf{R}}$, corresponding to a SIRNA or a  SIRCA is detailed in \cite{AdhikariAccess2}.  The dimensions of $\tilde{\mathbf{R}}$ are $((N-1)M+1)^2$-by-$((N-1)M+1)^2$ for SIRNA and $(MN+M)^2$-by-$(MN+M)^2$ for SIRCA.

\subsection{Application of MNM Algorithms to Planar Sparse Arrays}

\subsubsection{Direct MNM for Planar Arrays}

The MNM algorithm can be directly applied to the block-Toeplitz correlation estimate obtained in Section  \ref{sec:planarcorrelation}. By performing the eigenvalue decomposition of $\tilde{\mathbf{R}},$ the expression for the vector $\mathbf{d}$ can be evaluated. The resulting bivariate MNM pseudospectrum expression is given by
\begin{equation}
\label{mnm2}
P_{MN}(u_x,u_y)=\Bigl(|\tilde{\mathbf{v}}(u_x,u_y)^H\mathbf{d}|\Bigr)^{-2},
\end{equation}
where $\tilde{\mathbf{v}}(u_x,u_y)$ is the bivariate steering vector corresponding to direction cosines $u_x$ and $u_y$, as defined in Section  \ref{sec:planarcorrelation}. Note that for sparse arrays, the elements of $\tilde{\mathbf{v}}(u_x,u_y)$ do not correspond to physical sensor locations. For a SIRNA, $\tilde{\mathbf{v}}(u_x,u_y)$ is the Kronecker product between 
\begin{equation}
\label{wx}
\mathbf{w}_x= [e^{j\pi u_{x}0},e^{j\pi u_{x}1},e^{j\pi u_{x}2},\hdots , e^{j\pi u_{x}\tilde{m}}]^T
\end{equation}
and
\begin{equation}
\label{wy}
\mathbf{w}_y= [e^{j\pi u_{y}0},e^{j\pi u_{y}1},e^{j\pi u_{y}2},\hdots , e^{j\pi u_{y}\tilde{m}}]^T,
\end{equation}
with $\tilde{m}=(N-1)M$ for a SIRNA and $\tilde{m}=MN+M-1$ for a SIRCA.  The MUSIC pseudospectrum is given by 
\begin{equation}
\label{music2}
P_{MUSIC}(u_x,u_y)=\Bigl(\tilde{\mathbf{v}}(u_x,u_y)^H\tilde{\mathbf{E}}_n\tilde{\mathbf{v}}(u_x,u_y)\Bigr)^{-1},
\end{equation}
where $\tilde{\mathbf{E}}_n$ is the matrix comprised of noise eigenvectors of $\tilde{\mathbf{R}}$.

\subsubsection{Linear MNM for Planar Arrays}

One disadvantage of direct application of MNM or MUSIC to planar arrays is that the resulting spectrum is a bivariate function of $u_x$ and $u_y$, where $-1\leq u_x,u_y\leq 1$. Accurate DoA estimation requires searching for maxima of $P_{MN}(u_x,u_y)$ over a fine two-dimensional grid , which is computationally expensive.

We propose a linear MNM algorithm for sparse planar arrays. This method exploits the property of SIRNA and SIRCA that each row of sensors and each column of sensors are shifted copies of the underlying linear nested or coprime arrays. Treating every row $i$ of a SIRNA or SIRCA as a linear nested or coprime array, we can obtain the correlation matrix $\tilde{\mathbf{R}}_{x,i}$ by following the procedure to obtain $\mathbf{R}$ in \cite{AdhikariNAECON}. Then, the correlation matrix $\tilde{\mathbf{R}}_{x}$ for estimation of $u_x$ is obtained by averaging over all available $\tilde{\mathbf{R}}_{x,i}$. Thus, for SIRNA,
$\tilde{\mathbf{R}}_{x}=\sum_{i=1}^{N}\tilde{\mathbf{R}}_{x,i}/N$ and for SIRCA, $\tilde{\mathbf{R}}_{x}=\sum_{i=1}^{2M}\tilde{\mathbf{R}}_{x,i}/(2M)$. 

Similarly, treating every column $i$ of a SIRNA or SIRCA as a linear nested or coprime array, we can obtain the correlation matrix estimates $\tilde{\mathbf{R}}_{y,i}$. Then, $\tilde{\mathbf{R}}_{y}$ for estimation of $u_y$ is obtained by averaging over all available $\tilde{\mathbf{R}}_{y,i}$. The matrices for SIRNA and SIRCA are $\tilde{\mathbf{R}}_{y}=\sum_{i=1}^{N}\tilde{\mathbf{R}}_{y,i}/N$ and $\tilde{\mathbf{R}}_{y}=\sum_{i=1}^{2M}\tilde{\mathbf{R}}_{y,i}/(2M)$, respectively. Applying the MNM algorithm from Section  \ref{sec:correlation} using $\tilde{\mathbf{R}}_{x}$ and $\tilde{\mathbf{R}}_{y}$, we get the pseudospectrum estimates $P_{MN}(u_x)=(|\mathbf{w_x}^H\mathbf{d}|)^{-2}$ and $P_{MN}(u_y)=(|\mathbf{w_y}^H\mathbf{d}|)^{-2},$ where $\mathbf{w_x}$ and $\mathbf{w_y}$ are given in \eqref{wx} and \eqref{wy}. The $P$ peaks of $P_{MN}(u_x)$ are denoted by $\hat{u}_{x,i}$ and the $P$ peaks of $P_{MN}(u_y)$ are denoted by $\hat{u}_{y,i}$ for $i=1$, $2,$ $\hdots$, $P$. Since any $\hat{u}_{x,i}$ could be associated with any $\hat{u}_{y,i}$, there are $P^2$ possibilities for actual $(u_{x,i},u_{y,i})$ pairs. To find the right pairs, we evaluate \eqref{music2} at each possible pair. The $(\hat{u}_{x,i},\hat{u}_{y,i})$ pairs corresponding to the $P$ largest values of \eqref{music2} are the final estimates of the direction cosines.

\section{Results}
\label{sec:results}

\subsection{Linear Arrays Results}
\label{sec:linearresults}

\begin{figure}[h]
    \centering
    \includegraphics[width=0.45\textwidth,trim = 0cm 0cm 0cm 0cm]{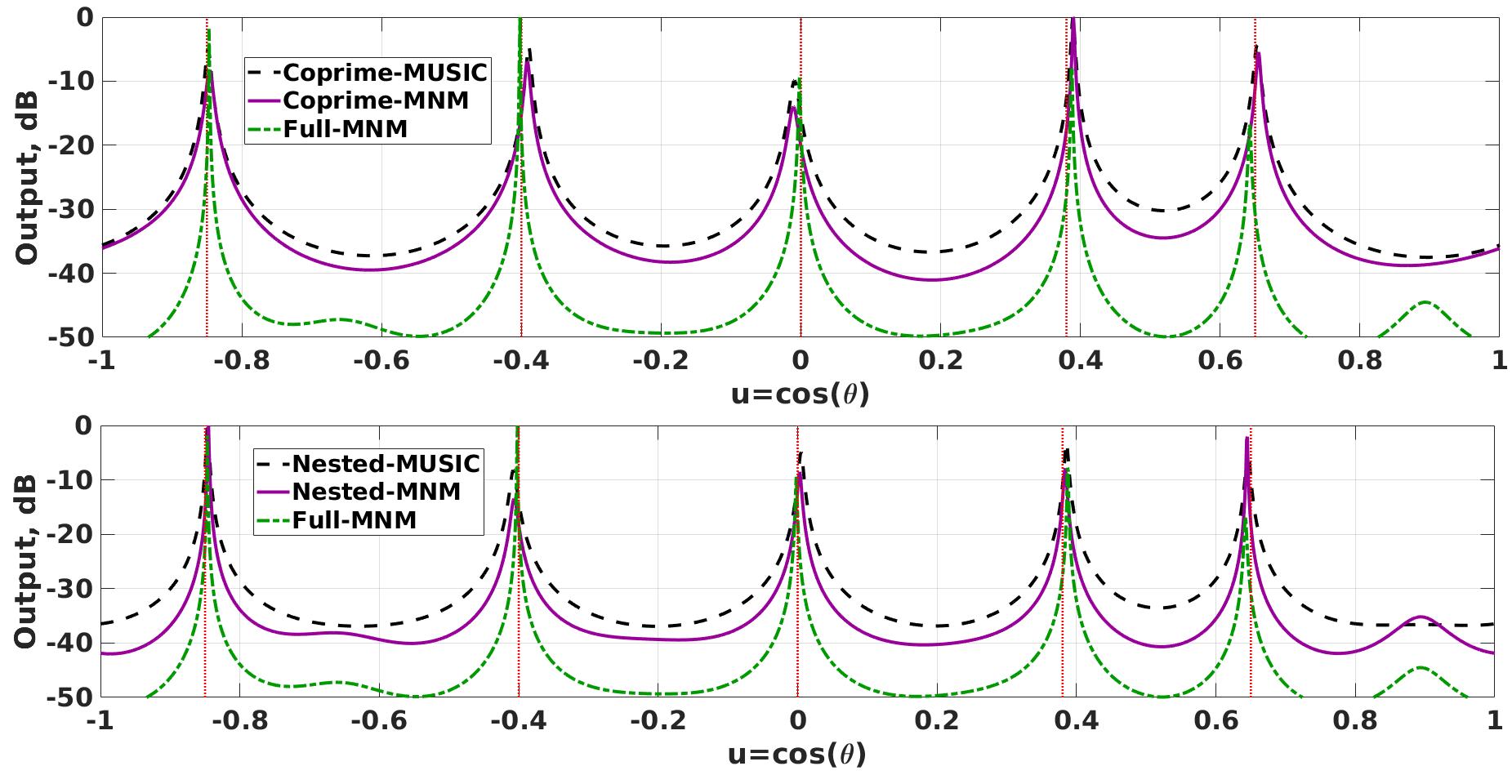}
    
    \caption{DoA estimation of five plane wave  sources. The number of snapshots is $100$, and the SNR is $0$ \rm{dB}. Top: Pseudospectra of the coprime array and the full ULA. Bottom: Pseudospectra of the nested array and the full ULA.\label{doafiglabel}}

\end{figure}

In this section, we use MNM to obtain spatial pseudospectra for coprime and nested arrays and compare the spectra with MUSIC outputs. We evaluate the performance metrics of MUSIC and MNM for various SNRs and numbers of snapshots.  For the linear case, we consider a coprime array, a nested array, and a full ULA with equal aperture as depicted in Fig. \ref{arrayslabel}. The parameters of the coprime array are $M_e=4,$ $N=2$, $N_e=4$, $M=3$, and the total number of sensors is $6$. The parameters of the nested array are $M_e=3,$ $N=1$, $N_e=4$, $M=3$, and the total number of sensors is $6$. The full ULA has $10$ sensors. The sizes of the correlation matrices obtained with the coprime, nested, and the full arrays are $7$-by-$7$, $8$-by-$8$, and $10$-by-$10$, respectively.  In the first example, we consider a scenario where there are five uncorrelated sources (i.e., $P=5$), the SNR is $0$~\rm{dB}, and the number of snapshots is $Q=100.$ The top panel of Fig. \ref{doafiglabel} illustrates the MUSIC spectrum (dashed line) of the coprime array, MNM spectrum (solid line) of the coprime array, and the MNM spectrum (dash-dot line) of the full ULA.  The bottom panel of Fig. \ref{doafiglabel} illustrates the corresponding outputs for the nested array. These plots show that both MNM and MUSIC  are able to identify the true source directions (dotted line) using the measurements received by the coprime and nested arrays. Since the full ULA has more sensors than the sparse arrays, its spectral estimates are better than the coprime and nested array outputs, as illustrated by sharper peaks and a lower noise floor. Comparing the MNM and MUSIC outputs for both coprime and nested arrays, we can infer that the MNM output is better than the MUSIC output. The MNM output exhibits narrower peaks at the true source locations and a lower noise floor.

Next, we evaluate two performance metrics, probability of resolution and RMSE, closely following the approach in \cite{VanTrees} to prove the superiority of MNM over MUSIC. We consider a scenario where there are two closely located uncorrelated plane wave signals impinging on the array from $\pm 0.5\Delta u_R,$ where $\Delta u_R$ is the half power bandwidth of the array. For a full array, $\Delta u_R=0.2165\times BW,$ where $BW$ is the null-to-null bandwidth and equals $4/L$ for a full array with $L$ sensors.  For coprime and nested arrays, $\Delta u_R$ is the half power bandwidth of the full array with an equal aperture. We measure the algorithms' capability to resolve the two signals. If an algorithm yields two distinct solutions and for each solution $\hat{u}_i,$ $|\hat{u}_i-u_i|\leq 0.5\Delta u_R,$ where $u_i$ is an actual source location, the two signals are considered resolved \cite{VanTrees}. The second metric we evaluate is the normalized RMSE in DoA estimation using the same three arrays from Fig. \ref{arrayslabel}. The RMSE is normalized by dividing it by $BW$ \cite{VanTrees}.  We execute $1,000$ trials to evaluate the probabilities of resolution and RMSEs. 

\begin{figure}[h]
    \centering
    \includegraphics[width=0.45\textwidth,trim = 0cm 13cm 0cm 0cm]{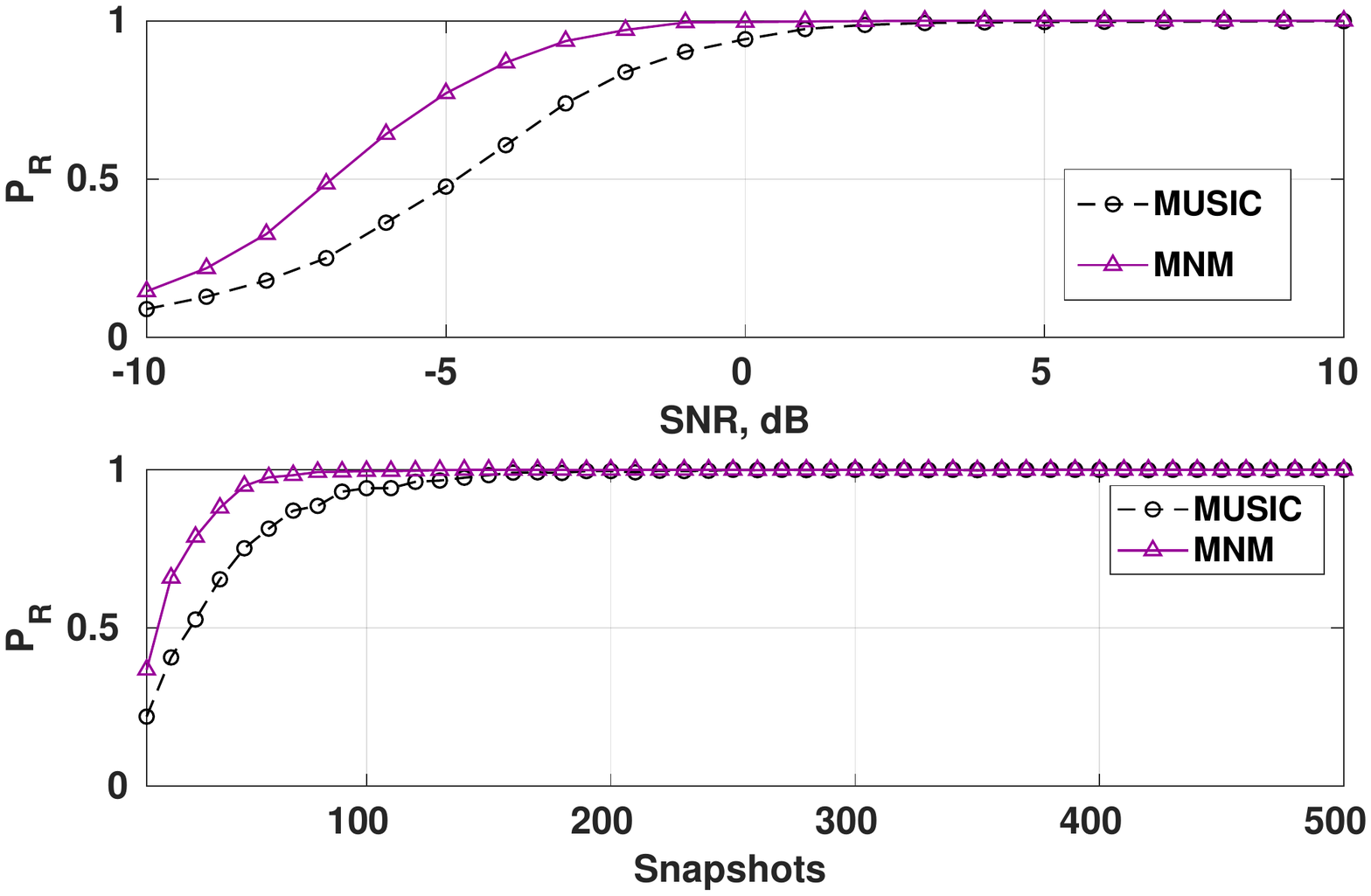}
    
    \caption{Comparison of probabilities of resolution for MUSIC and MNM for a nested array with $M_e=3$, $N_e=4$, $N=1$, and $M=3.$ Top: The number of snapshots is $100$.  Bottom: SNR is $0$ \rm{dB}.\label{prlabelnested}}

\end{figure}

\begin{figure}[h]
    \centering
    \includegraphics[width=0.45\textwidth,trim = 0cm 13cm 0cm 0cm]{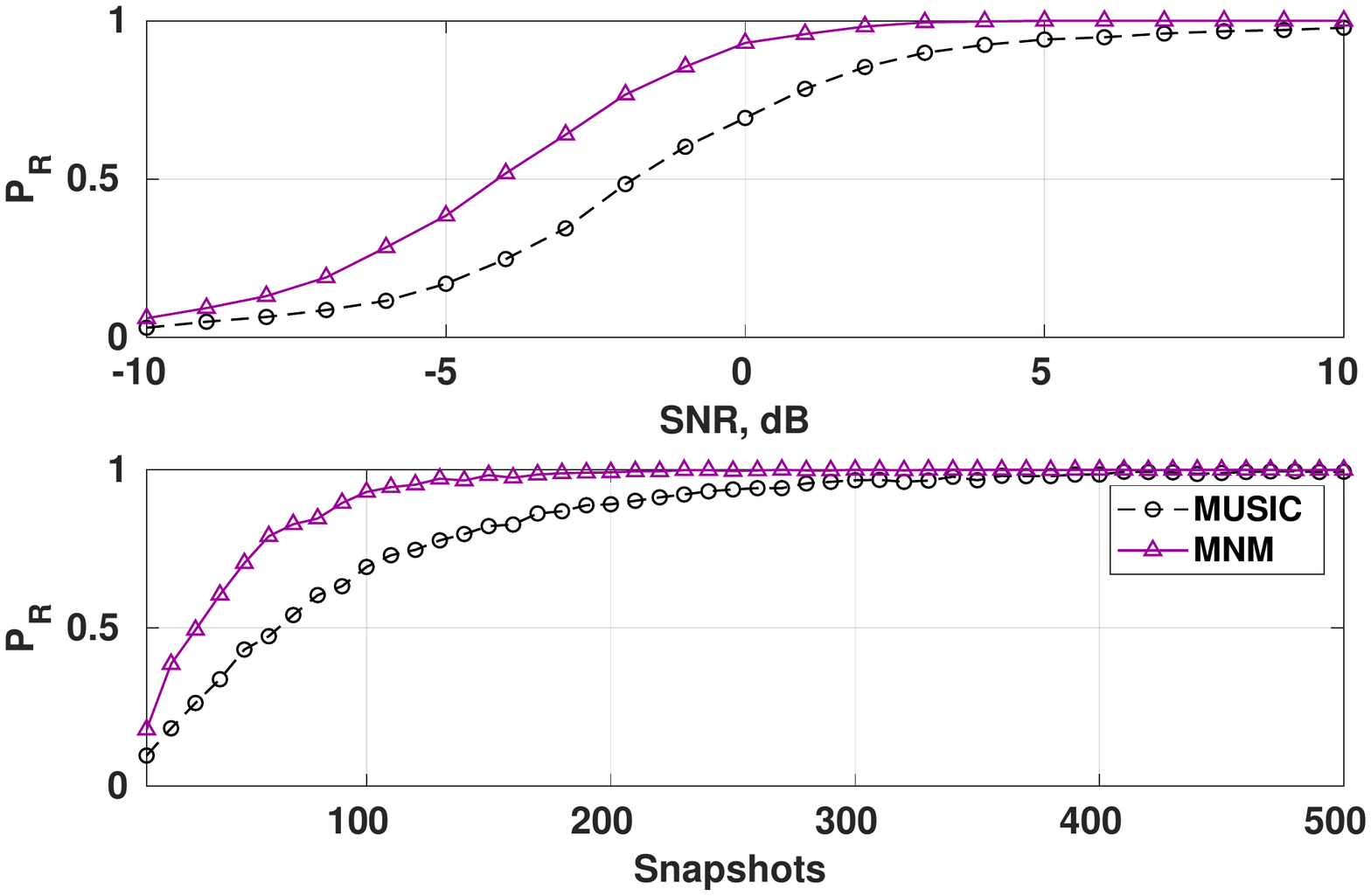}
    
    \caption{Comparison of probabilities of resolution for MUSIC and MNM for a coprime array with $M_e=4$, $N_e=4$, $N=2$, and $M=3.$ Top: The number of snapshots is $100$.  Bottom: SNR is $0$ \rm{dB}.\label{prlabelcoprime}}

\end{figure}
\begin{figure}[t]
    \centering
    \includegraphics[width=0.45\textwidth,trim = 0cm 13cm 0cm 0cm]{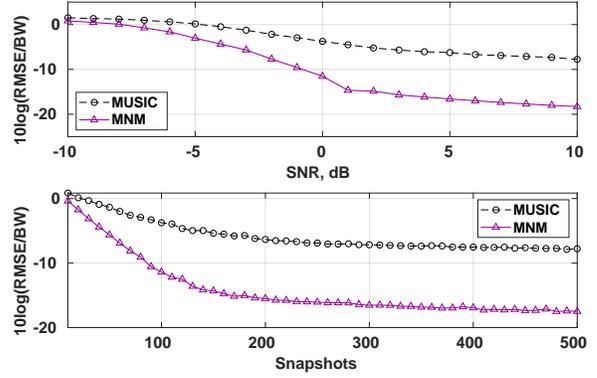}
    
    \caption{Comparison of RMSEs for MUSIC and MNM for a nested array with $M_e=3$, $N_e=4$, $N=1$, and $M=3.$ Top: The number of snapshots is $100$.  Bottom: SNR is $0$ \rm{dB}.\label{mselabelnested}}

\end{figure}

\begin{figure}[t]
    \centering
    \includegraphics[width=0.45\textwidth,trim = 0cm 13cm 0cm 0cm]{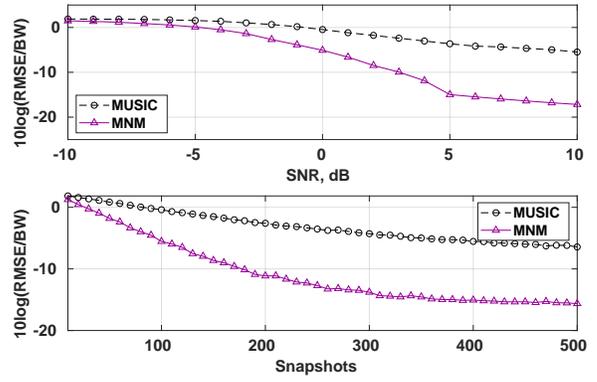}
    
    \caption{Comparison of RMSEs for MUSIC and MNM for a coprime array with $M_e=4$, $N_e=4$, $N=2$, and $M=3.$ Top: The number of snapshots is $100$.  Bottom: SNR is $0$ \rm{dB}.\label{mselabelcoprime}}

\end{figure}

Fig. \ref{prlabelnested} compares the probabilities of resolution for the nested array from Fig. \ref{arrayslabel} using MNM and MUSIC. The top panel compares them  over a range of SNR values at a fixed number of $100$ snapshots, and the bottom panel compares them  over a range of numbers of snapshots at a fixed $0$ \rm{dB} SNR. Fig. \ref{prlabelcoprime} compares the corresponding probabilities of resolution for the coprime array from Fig. \ref{arrayslabel}. The figures confirm that the MNM algorithm offers higher probabilities of resolution than the MUSIC algorithm for both sparse arrays over the ranges of SNRs and numbers of snapshots. The RMSE plots for the coprime and nested arrays for the two algorithms are plotted in Fig. \ref{mselabelnested} and Fig. \ref{mselabelcoprime}. The top panels of the figures illustrate the RMSEs for a range of SNRs when the number of snapshots is fixed at $100$, and the bottom panels illustrate the RMSEs for a range of snapshots when the SNR is fixed at $0$ \rm{dB}.  The MNM algorithm demonstrates better accuracy than MUSIC over the entire ranges of SNRs and numbers of snapshots for both coprime and nested arrays.

\subsection{Linear MNM Results for the Planar Arrays}
\label{sec:resultsindirect}

\begin{figure}[t]
    \centering
    \includegraphics[width=0.4\textwidth,trim = 0cm 13cm 0cm 0cm]{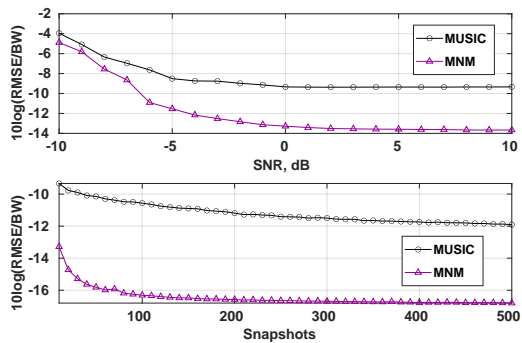}
    
    \caption{Comparison of RMSEs for MUSIC and MNM for a SIRNA with $M=3$ and $N=4$. Top: The number of snapshots is $15$.  Bottom: The SNR is $0$ \rm{dB}.\label{mselinear2dnested}}

\end{figure}

To compare the MNM and MUSIC algorithms for the planar arrays in Fig. \ref{labelsir}, we consider a scenario with two closely located plane wave signals, each with $0$ \rm{dB} SNR and located at $(u_x,u_y)$ values of $(0.297,0.46)$ and $(0,-0.094)$. Direct MNM requires searching for the maxima in two-dimensional pseudospectra to find the signal directions, which is computationally expensive. Thus, we recommend using the linear MNM and MUSIC algorithms for these planar arrays. We use the planar arrays in Fig. \ref{labelsir} again. The RMSE plots corresponding to $0$ \rm{dB} SNR sources at $(u_x,u_y)=(0.297,0.46)$ and $(u_x,u_y)=(0,-0.094)$ are illustrated in Fig. \ref{mselinear2dnested} for the SIRNA and Fig. \ref{mselinear2dcoprime} for the SIRCA. The top panels of the figures have a fixed number of $10$ snapshots and the SNR varies from $-10$ \rm{dB} to $10$ \rm{dB}. The bottom panels have a fixed SNR of $0$ \rm{dB}, and the number of snapshots varies from $10$ to $500$. The figures show that the MNM algorithm bests the MUSIC algorithm for planar sparse arrays as well.

\section{Conclusion}
\label{sec:conclusion}

In this work, we compared two eigenanalysis-based methods for DoA estimation of uncorrelated WSS plane wave signals in spatially white Gaussian noise using linear and planar sparse arrays. We demonstrated higher accuracy and higher probability of resolution of the MNM algorithm than the MUSIC algorithm for linear sparse arrays. Comparison of the pseudospectra of MNM and MUSIC for both linear coprime and nested arrays show that MNM exhibits narrower peaks at true source directions and a lower noise floor.

For two types of planar sparse arrays, we presented two variants of the MNM algorithm and compared them with the equivalent variants of the MUSIC algorithm. We showed that linear MNM exhibits higher accuracy than the MUSIC algorithm for the planar arrays as well. 
  The proposed algorithm is used for DoA estimation of uncorrelated WSS plane wave signals in spatially white noise.  Our results show that MNM has lower mean squared error than MUSIC in DoA estimation for both coprime and nested arrays.

\begin{figure}[t]
\centering
\includegraphics[width=0.4\textwidth,trim = 0cm 13cm 0cm 0cm]{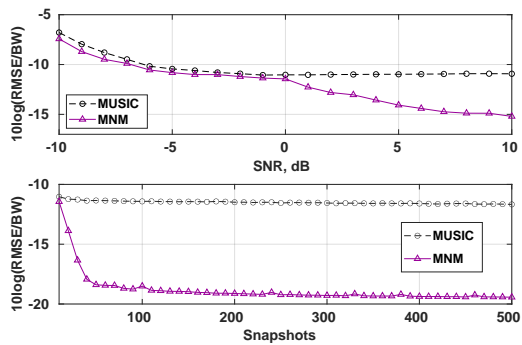}

\caption{Comparison of RMSEs for MUSIC and MNM for a SIRCA with $M=2$. Top: The number of snapshots is $15$.  Bottom: The SNR is $0$ \rm{dB}.\label{mselinear2dcoprime}}
\label{mselinear2dcoprime}
\end{figure}
\bibliographystyle{IEEEtran}
\bibliography{IEEEabrv,MinNormReferences3}

\end{document}